\documentclass[twocolumn,superscriptaddress,aps,prd,preprintnumbers,amsmath,amssymb,sort&compress,nofootinbib]{revtex4-1}

\usepackage[utf8]{inputenc}
\usepackage[T1]{fontenc}

\usepackage{amsfonts,amsmath,amssymb,bm}
\usepackage{comment}
\usepackage{ifpdf}
\usepackage{slashed}
\usepackage{color}
\usepackage[mathscr]{eucal}
\usepackage[utf8]{inputenc}

\usepackage{cancel}

\definecolor{red}{rgb}{1,0,0}
\definecolor{darkred}{rgb}{0.6,0,0}
\definecolor{darkgreen}{rgb}{0.992447,0.623778,0.034597}
\definecolor{ppink}{rgb}{1,0.4,0.4}
\definecolor{bblue}{rgb}{0.284602,0.317763,0.963947}

  \usepackage{grffile}     
  \usepackage{graphicx}     
  \usepackage[bookmarksopen,colorlinks=true,linkcolor=bblue,citecolor=bblue,    urlcolor=ppink]{hyperref}

\usepackage{verbatim}
\usepackage{rotate}
\usepackage{aas_macros}

\usepackage{savesym}
\savesymbol{tablenum}
\usepackage{siunitx}
\restoresymbol{SIX}{tablenum}

\definecolor{RedWine}{rgb}{0.743,0,0}
\definecolor{GrassGreen}{rgb}{0.125,0.75,0.125}
\definecolor{RoyalBlue}{rgb}{0.25,0.41,0.88}

\newcommand{\be}{\begin{equation}}
\newcommand{\ee}{\end{equation}}
\newcommand{\bea}{\begin{eqnarray}}
\newcommand{\eea}{\end{eqnarray}}
\def\ba#1\ea{\begin{align}#1\end{align}}

\def\({\left(}
\def\){\right)}
\def\<{\left\langle}
\def\>{\right\rangle}

\newcommand{\refeq}[1]{\cref{eq:#1}}
\newcommand{\refeqs}[2]{\crefrange{eq:#1}{eq:#2}}
\newcommand{\reffig}[1]{\cref{fig:#1}}
\newcommand{\reffigs}[2]{\crefrange{fig:#1}{fig:#2}}
\newcommand{\refsec}[1]{\cref{sec:#1}}

\newcommand{\vs}{\nonumber\\}

\def\vr{{\bm{r}}}
\def\vx{{\bm{x}}}
\def\vv{{\bm{v}}}

\def\vk{{\bm{k}}}

\def\vq{{\bm{q}}}

\def\nhat{{\hat{\bm{n}}}}
\def\khat{{\hat{\bm{k}}}}

\def\rhat{{\hat{\bm{r}}}}

\def\zhat{{\hat{\bm{z}}}}

\DeclareMathOperator{\erf}{erf}

\DeclareSIUnit \parsec {pc}
\DeclareSIUnit \h {\text{$h$}}
\DeclareSIUnit \year {yr}
\DeclareSIUnit \solarmass {M_\odot}
\DeclareSIUnit \Mpc {\mega\parsec}

\def\dd{\mathrm{d}}

\def\bdelta{\bar{\delta}}

\def\myapp#1#2{%
  \mathrel{%
    \setbox0=\hbox{$#1\sim$}%
    \setbox2=\hbox{%
      \rlap{\hbox{$#1\propto$}}%
      \lower1.1\ht0\box0%
    }%
    \raise0.25\ht2\box2%
  }%
}
\def\approxpropto{\mathpalette\myapp\relax}

\newcommand{\incgraph}[2][0.49]{\includegraphics[width=#1\textwidth]{#2}}

\usepackage[capitalise]{cleveref}

\begin{document}

\title{%
Nonlinear Redshift-Space Distortions in the Harmonic-space Galaxy Power Spectrum
}%
\author{Henry S. \surname{Grasshorn Gebhardt}}
\email{henry.s.gebhardt@jpl.nasa.gov}
\altaffiliation{now NASA Postdoctoral Program Fellow located at Jet Propulsion Laboratory, 4800 Oak Grove Drive, Pasadena, CA 91109}
\affiliation{Department of Astronomy and Astrophysics and
    Institute for Gravitation and the Cosmos, \\
    The Pennsylvania State University, University Park, PA 16802, USA
}
\author{Donghui Jeong}
\email{djeong@psu.edu}
\affiliation{Department of Astronomy and Astrophysics and
    Institute for Gravitation and the Cosmos, \\
    The Pennsylvania State University, University Park, PA 16802, USA
}
\begin{abstract}
Future high spectroscopic resolution galaxy surveys will observe galaxies with
nearly full-sky footprints. Modeling the galaxy clustering for these surveys,
therefore, must include the wide-angle effect with narrow redshift binning. In
particular, when the redshift-bin size is comparable to the typical peculiar
velocity field, the nonlinear redshift-space distortion (RSD) effect becomes
important. A naive projection of the Fourier-space RSD model to spherical
harmonic space leads to diverging expressions. In this paper we present a
general formalism of projecting the higher-order RSD terms into spherical
harmonic space. We show that the nonlinear RSD effect, including the
fingers-of-God (FoG), can be entirely attributed to a modification of the
radial window function. We find that while linear RSD enhances the
harmonic-space power spectrum, unlike the three-dimensional case, the
enhancement decreases on small angular-scales. The fingers-of-God suppress the
angular power spectrum on \emph{all} transverse scales if the bin size is
smaller than $\Delta r \lesssim \pi \sigma_u$; for example, the radial bin
sizes corresponding to a spectral resolution $R=\lambda/\Delta \lambda$ of a
few hundred satisfy the condition. We also provide the flat-sky approximation
which reproduces the full calculation to sub-percent accuracy.
\end{abstract}

\maketitle

\section{Introduction}
Future galaxy redshift surveys such as 
Euclid \citep{Amendola+:2018LRR....21....2A}, 
DESI (Dark Energy Survey Instrument) \citep{DESICollaboration:2016arXiv161100036D}, 
and 
SPHEREx (Spectro-Photometer for the History of the Universe, Epoch of
Reionization, and Ices Explorer) \citep{Dore+:2014arXiv1412.4872D} 
plan to cover nearly full-sky footprints. With the line of sight changing
significantly over the survey footprint,
it is clear that full exploitation of the cosmological 
information in these surveys requires analysis beyond the 
usual plane-parallel (or distant observer) approximation that assumes a
single line of sight throughout the survey volume.

The galaxies' peculiar velocities in the direction of the line of sight
complicate the analysis of wide, nearly full-sky surveys; the peculiar velocities
contribute to the observed redshift in addition to the Hubble flow, causing
an offset between the actual distances and those inferred from
observed redshifts.
This phenomenon is called redshift-space distortion (RSD), and we have 
the theoretical templates for modeling RSD in the following two regimes.

In the linear regime, or on large scales, galaxies' peculiar velocities 
are determined by the linear growth of the cosmic density field. That is, 
the growth of the cosmic density field derives coherent inflows to the
overdensity and outflows from the underdensity. 
Adopting the plane-parallel approximation, 
Ref.~\cite{Kaiser:1987MNRAS.227....1K}
has first derived the expression for the observed galaxy power spectrum 
with RSD, and Ref.~\cite{hamilton:1992} has found the corresponding expression 
for the galaxy 2PCF (two-point correlation function) in configuration space.
For wide-angle galaxy surveys, Refs.~\cite{Fisher+:1994MNRAS.266..219F,Heavens/Taylor:1995,Hamilton/Culhane:1995,Zaroubi/Hoffman:1996,Szalay/Matsubara/Landy:1997,Matsubara:1999,Szapudi:2004,Papai/Szapudi:2008}
have extended the formulae to obtain the expressions for the linear 
two-point correlation functions with RSD: 
$\xi(r_1,r_2,\theta)$ in configuration space, 
$C_\ell(r_1,r_2)$ in spherical harmonic space, or
$C_\ell(k_1,k_2)$ in spherical Fourier-Bessel space.

In the highly nonlinear regime, or on small scales, where galaxies predominantly reside in gravitationally bounded structures such as galaxy
clusters, the random peculiar velocities of galaxies 
\citep{Jackson:1972MNRAS.156P...1J} manifest themselves in redshift
space by stretching the galaxy clusters. This effect creates an 
observational illusion that artificially puts the observer in a
special location as if all galaxy clusters were pointing at her:
\citet{Tully+:1978IAUS...79...31T} called these the \emph{Fingers of God}
(FoG). Caused by the random velocities in virialized clusters,
one can model the elongated {\it fingers} by convolving the shape of the 
galaxy clusters with the line-of-sight velocity distribution function \cite{Peacock/Dodds:1994,Heavens/Taylor:1995}. In particular, convolving 
the 2PCF in real space with the LoSPVDF 
(line-of-sight pair-wise velocity distribution function) 
yields the 2PCF in redshift space. 
The two widely-used phenomenological models for the LoSPVDF in literature
are the Gaussian \cite{Peacock/Dodds:1994} pdf 
(probability distribution function) 
and the exponential \cite{Peebles:1976} pdf.

Thus far, the use of the wide-angle formula for the analysis of
galaxy surveys has been limited to the following few publications.
Refs.~\cite{Heavens/Taylor:1995,Ballinger/etal:1995,Tadros/etal:1999,Percival/etal:2004} have applied the spherical Fourier-Bessel 
basis formula for the clustering analysis of, respectively,
the 1.2-Jy survey \cite{Fisher/etal:1995}, PSCz surveys 
\cite{Sauders/etal:2000} using IRAS (The Infrared Astronomical Satellite), and 
2dFGRS (2dF Galaxy Redshift Survey) \cite{Colless/etal:2001,2dFGRS:2003}.
Focusing on large scales, $k\lesssim\SI{0.15}{\h\per\mega\parsec}$, and on measuring
the RSD parameter $\beta=f/b_1$, the ratio between the linear growth 
rate ($f\equiv d\ln D/d\ln a$ where $D(a)$ is the linear growth factor) and 
the linear bias parameter $b_1$, they find that the FoG effect hardly 
changes the measurement of the RSD parameter. In these analyses, 
the LoSPVDF is often assumed to follow a Gaussian pdf, for which the 
FoG effect merely rescales the redshift uncertainties.
More recently, Ref.~\cite{Beutler/etal:2019} has applied the wide-angle 
formula in configuration space to the BOSS DR12 
\cite{Ross/etal:2016,Beutler/etal:2016} dataset.
The harmonic space formula has been used to analyze the galaxy clustering 
tomography in \cite{Fisher+:1994MNRAS.266..219F,Balaguera-Antol/etal:2018,Loureiro+:2019MNRAS.485..326L}, for example.

For the current generation of galaxy surveys, the systematic effects of the
plane-parallel (or distant-observer) approximation are negligibly small
\cite{Samushia/etal:2012, Yoo/Seljak:2015}. Furthermore,
Ref.~\cite{Yoo/Seljak:2015} has also shown that, even for future surveys such
as DESI and Euclid, one can reduce the wide-angle effect in the 2PCF multipoles
$\xi_\ell(r)$ and $P_\ell(k)$ by employing the local line-of-sight 
estimator \cite{Percival:2018}.

We stress, however, that such an approximation is only possible for the
auto-correlation analyses of galaxies. The cross-correlation between galaxy
distributions at different redshifts or between galaxies and various full-sky
maps (for example, CMB anisotropies, weak gravitational lensing map) must be
analyzed by using the spherical bases, either in configuration (angular) space or
spherical harmonic space. Otherwise, mimicking the angular cross-correlation
requires a clumsy coordinate transformation, as we have done in
Ref.~\cite{GrasshornGebhardt/etal:2019}. The spherical bases are also natural to
incorporate the redshift evolution of physical quantities such as the galaxy
bias, galaxy number density, and linear growth rates, which are kept constant in
the usual plane-parallel analysis. In the companion
paper (Ref.~\cite{paper3}), we shall show that including the radial evolution
of these quantities can improve the accuracy of the
geometrical measurement of the Hubble expansion rate and the angular diameter
distance.

In this paper, we shall focus on the angular 2PCF in harmonic space
$C_\ell(r_1,r_2)$, which can be thought of as, 
for large-scale spectroscopic surveys, 
a fine-radial-binning version of the traditional 2D tomography analysis.
Ref.~\cite{Asorey/etal:2012} shows that a fine redshift binning with 
\be
\frac{\Delta z}{z} 
\lesssim 
\frac{\pi H}{zk_{\rm max}}
\simeq 
0.008 \left(\frac{0.2~h/{\rm Mpc}}{k_{\rm max}}\right)\,,
\label{eq:deltaz}
\ee
is required for the angular-basis analysis to recover the full information 
in the galaxy 2PCF. Here, $k_H=aH$ is the comoving horizon wavenumber, and 
the approximation holds for $1\lesssim z\lesssim5$.

The future large-scale spectroscopic galaxy surveys with high galaxy 
sample densities make the angular clustering analysis possible with such a 
narrow radial binning. For example, with the designed sensitivity, 
the Euclid satellite can observe 50 million galaxies in the redshift range 
$0.9<z<1.8$ \citep{Amendola+:2018LRR....21....2A}, 
which translates to about a quarter-million objects in a redshift bin of 
size $\Delta z/z\sim0.005$.

One of the challenges in analyzing the galaxy surveys in harmonic space is
that the calculation of the angular power spectra $C_\ell(r_1,r_2)$ involves
highly oscillating integrals of the form
\be
C_{\ell\ell'}^{(n,n',\alpha)}(r_1,r_2)
=
\frac{2}{\pi} \int\dd k \, j_\ell^{(n)}(kr_1) \, j_{\ell'}^{(n')}(kr_2)
\, k^{\alpha}
\, P(k)\,,
\label{eq:Cll}
\ee
where $j_\ell^{(n)}(kr)$ is the $n$-th derivative of the spherical Bessel
function, and $P(k)$ is the power spectrum. For the full analysis, one needs
to evaluate \refeq{Cll} for all combinations of $r_1$ and $r_2$; for the
Euclid example above there are about \num{16000} different combinations of
$r_1$ and $r_2$. The recent development of the 2-FAST algorithm
\citep{GrasshornGebhardt+:2018PhRvD..97b3504G} (see also
\citep{Assassi+:2017JCAP...11..054A}) resolves this issue by evaluating \refeq{Cll} fast and accurate. The key ideas are the FFTlog-based transformation that converts the integration to the
hypergeometric function ${}_2F_1$ and a stable recurrence relation that
accelerates the evaluation of ${}_2F_1$.

Another challenge, which we address in this paper, is the nonlinear RSD
effect that becomes significant in $C_{\ell}(r_1,r_2)$ with a fine radial
binning satisfying the condition in \refeq{deltaz}. The importance of the RSD
effect shall become apparent in the examples in later sections. However, it is
simple to understand: At redshift $z\sim1$, the redshift bin width $\Delta
z\simeq 0.005$ corresponds to a peculiar velocity of
\SI{750}{\kilo\meter\per\second}. That is the same order of magnitude as the
typical peculiar velocities of galaxies in the galaxy groups or clusters.
Therefore, the peculiar velocities move galaxies from one radial bin to
another, and the FoG effect is in action for $C_{\ell}(r_1,r_2)$ with small
radial binning.

Of course, when the FoG effect is important, the modeling must also include
the nonlinear Kaiser effect
\cite{Scoccimarro:2004,taruya/nishimichi/saito:2010} that captures the
nonlinearities on intermediate scales.
Ref.~\citep{2020PhRvD.101d3530J} compares several non-linear
extensions using the flat-sky approximation.
For modeling the nonlinear Kaiser
effect without the plane-parallel approximation, Ref.~\citep{Shaw/Lewis:2008}
works out the wide-angle formalism including the nonlinear RSD transformation
by assuming that the velocity field follows Gaussian statistics, and recent
studies in Ref.~\citep{Castorina/White:2018,Taruya/etal:2019} have developed
the formalism in quasi-linear scales and Gaussian FoG by using the
Zel'dovich approximation.

While the wide-angle formula corresponding to the full nonlinear Kaiser
effect in Refs.~\cite{Scoccimarro:2004,taruya/nishimichi/saito:2010} is
desirable to fully exploit the galaxy power spectrum of large surveys, there is a
more straightforward, but perhaps more urgent, problem that
arises when extracting the Baryon Acoustic Oscillations (BAO) from
$C_\ell(r_1,r_2)$ statistics.
The details of the BAO analysis will be presented in a forthcoming
paper. Here, we content ourselves with setting the context for the problem of
convergence.
Given that the CMB measurement fixes
the sound horizon scale at the baryon-decoupling epoch, BAO is a standard
ruler used by all dark-energy driven galaxy surveys (see
\cite{weinberg/DEreview:2013} for a review). Because the late-time
nonlinearities do not shift the location of the peaks in the real space 2PCF
\cite{seo/etal:2010}, the standard procedure of modeling the BAO in Fourier
space after the reconstruction \cite{padmanabhan/etal:2012} is to model the
anisotropic damping due to the bulk flow \cite{seo/eisenstein:2007} by
introducing an anisotropic smoothing function 
\be
P_{\rm BAO, nl}
=
P_{\rm BAO, lin}\,
e^{-k^2\left(\mu^2\Sigma_\parallel^2 + (1-\mu^2)\Sigma_\perp^2\right)/2}\,,
\ee
with $\Sigma_\parallel$ and $\Sigma_\perp$ being r.m.s.\ 
displacements in Lagrangian space, respectively, along the line-of-sight 
and perpendicular directions \cite{seo/etal:2016}, and 
$\mu\equiv \khat\cdot\nhat$.
Even before reconstruction, we can also extract the phase of the BAO 
in redshift space by modeling or subtracting the no-wiggle part that can be 
captured by a polynomial expansion of the form $k^a\mu^b$ 
\cite{shoji/etal:2009,GrasshornGebhardt/etal:2019}.

How do we calculate the harmonic space expression corresponding to these
treatments of nonlinearities in the Fourier space? The problem occurs when one
tries to obtain the perturbative solution by Taylor-expanding the exponential
function because the projection integral in \refeq{Cll} does not converge for
all powers $\alpha\ge5$. The solution that we suggest is to extend the convolution integral that, for example, Ref.~\cite{Heavens/Taylor:1995} has adopted to model the FoG effect. Including the polynomial nonlinear Kaiser contributions, one can define new convolution kernels.
In this case, the calculation of the harmonic space $C_\ell(r_1,r_2)$ boils down to three convolutions: 
two from redshift-bin window functions and one from the nonlinear Kaiser
effect. Note, however, that we can further reduce the number of convolutions 
to two by using integration by parts. This method is similar to that of 
Refs. \cite{Assassi+:2017JCAP...11..054A,Schoneberg/etal:2018} simplifying the
linear Kaiser effect calculation. The net effect is distributing the 
nonlinear Kaiser effect to re-define the window function; by using these 
new window functions, we only need to evaluate the convolution twice for 
each calculation of $C_\ell(r_1,r_2)$. 
The main goal of this paper is to study this novel method and verify
it by comparing the predictions to the simulations 
\citep{Agrawal+:2017JCAP...10..003A}.

For the calculations in \refsec{flatsky} and \refsec{comparisons},
we use a flat $\Lambda$CDM \emph{Planck} cosmological parameters
\citep{PlanckCollaboration:2018arXiv180706205P,PlanckCollaboration:2018arXiv180706209P}
with the fiducial values 
$\Omega_{\Lambda}=0.69179$,
$\Omega_{b0}h^2=0.022307$,
$\Omega_{c0}h^2=0.11865$,
$\Omega_{\nu0}h^2=0.000638$,
$h=0.6778$, and $n_s=0.9672$. 
We calculate the linear power spectrum $P(k)$ using the
\citet{Eisenstein+:1998ApJ...496..605E} fitting formula.
With this cosmological parameters, the linear growth rates are 
$f=0.541$ and $0.706$, respectively, for the comoving radial distances of
$r_0=\frac12(r_1+r_2)=\SI{100}{\per\h\mega\parsec}$ and
$r_0=\SI{1000}{\per\h\mega\parsec}$.
We set the linear galaxy bias $b=1$. 

This paper is organized as follows. 
In \refsec{FoG}, we summarize the problem of calculating the 
nonlinear Kaiser effect perturbatively for the harmonic space power spectrum.
In \refsec{wide-angle} we derive the method for general nonlinear 
Kaiser terms, and in \refsec{winfn} we work out an example of the FoG effect.
Finally, in \refsec{flatsky} and \refsec{comparisons},
we compare the results with, respectively, the flat-sky 
and log-normal simulations. We conclude in \refsec{conclusion}.

\section{Diverging integrals in the angular power spectrum of galaxies}
\label{sec:FoG}
In this section, we illustrate the difficulty of calculating the harmonic
space power spectrum $C_\ell(r_1,r_2)$ with a
perturbative modeling of the nonlinear Kaiser effect, for example, as shown in 
Ref.~\cite{Desjacques/etal:2018}.
We use the FoG effect as an example, but the same applies to the general 
nonlinear expression beyond the linear Kaiser effect. In Fourier space, 
the observed density contrast $\delta_g^\mathrm{RSD}(\vk)$ is expressed 
in terms of the real-space density contrast by
\ba
\label{eq:delta_g}
\delta_g^\mathrm{RSD}(\vk)
&= \widetilde A_\mathrm{RSD}(\mu,k\mu) \, \delta_g^\mathrm{real}(\vk)\,,
\ea
where $\mu\equiv\khat\cdot\nhat$ with the line of sight ($\nhat$), and we 
break down the operator 
$\widetilde{A}_{\rm RSD}$ into the linear Kaiser part and and the 
nonlinear part $\widetilde{A}_{\rm nl}$:
\ba
\label{eq:Arsd}
\widetilde A_\mathrm{RSD}(\mu,k\mu)
&=
\(1 + \beta\mu^2\) \widetilde A_\mathrm{nl}(\mu,k\mu)\,.
\ea
Here, $\beta\equiv b/f$ with the linear galaxy bias $b$ and 
the linear growth rate $f\equiv d\ln D/d\ln a$.

As an illustrative example, we consider the following three functional forms
for the nonlinear operator,
\ba
\label{eq:Afog_a}
\widetilde A^a_\mathrm{nl}(k\mu)
&= e^{-\frac12 \sigma_u^2 k^2\mu^2}\,,
\\
\label{eq:Afog_b}
\widetilde A^b_\mathrm{nl}(k\mu)
&= \frac{1}{1+\frac12\sigma_u^2k^2\mu^2}\,,
\\
\label{eq:Afog_c}
\widetilde A^c_\mathrm{nl}(k\mu)
&= \frac{1}{\sqrt{1+\sigma_u^2k^2\mu^2}}\,,
\ea
where $\sigma_u^2$ is the one dimensional velocity dispersion in units of length,
\ba
\sigma_u
&= \SI{1}{Mpc/h}
\(\frac{\sigma_v}{100\,{\rm km/s}}\)
\(\frac{1+z}{H(z)/(100\,h\,{\rm km/s/Mpc})}\)
\vs
&\simeq
\SI{0.88}{Mpc/h}
\(\frac{\sigma_v}{100\,{\rm km/s}}\)
\(\frac{1+z}{4}\)^{-0.4}\,,
\ea
where $\sigma_v$ is in \si{\kilo\meter\per\second} and the last line
holds approximately for $1<z<5$.
Note that the tilde attached to the $\widetilde A_{\rm nl}$ operators signifies that they 
are defined in Fourier space.
The three forms in \refeqs{Afog_a}{Afog_c} correspond to three models for the
FoG, a Gaussian suppression \citep{Peacock/Dodds:1994}, a Lorentzian
suppression 
\citep{Peebles:1976,Ballinger/Peacock/Heavens:1996,Taylor+:2001MNRAS.328.1027T,Percival/etal:2004}, and a square-root Lorentzian suppression
\cite{Percival/etal:2004}.
Refs. \cite{Ratcliffe+:1998MNRAS.296..191R,Landy:2002ApJ...567L...1L}
find that a Lorentzian FoG is in better agreement with measurements. 

Now, let us consider the harmonic-space transformation of \refeq{delta_g}:
\be
\delta_g^{\rm RSD}(r\nhat)
=
\sum_{\ell m} \delta_{\ell m,g}^{\rm RSD}(r) Y_{\ell m}(\nhat)\,,
\ee
with the harmonic-space coefficients
\ba
\delta_{\ell m,g}^{\rm RSD}(r)
=&
\int d\Omega \, Y_{\ell m}^*(\nhat)
\vs&\times
\int \frac{d^3k}{(2\pi)^3} \, e^{ir\vk\cdot\nhat}
\,\widetilde{A}_{\rm RSD}(\mu,k\mu) \, \delta_g^{\rm real}(\vk)\,.
\label{eq:almr0}
\ea
We then may write a generic RSD term in perturbation theory as an expansion 
in $k^n\mu^p$, i.e.\ 
\ba
\label{eq:series-expansion}
\widetilde A_\mathrm{RSD}(\mu,k\mu) = \sum_{np} d_{np}\,k^n\mu^p\,,
\ea
with some coefficients $d_{np}$ which are proportional to $\sigma_u^n$ for the 
case of FoG terms listed in \refeqs{Afog_a}{Afog_c}.
The angular power spectrum using the perturbative expansion is given as
\ba
&
\left<
\delta_{\ell m,g}^{\rm RSD^*}(r)
\delta_{\ell' m',g}^{\rm RSD}(r')
\right>
\vs
=&
\int d\Omega_{\nhat} Y_{\ell m}(\nhat)
\int d\Omega_{\nhat'} Y_{\ell' m'}^*(\nhat')
\vs
&\times
\int\frac{d^3k}{(2\pi)^3} e^{i\vk\cdot(r'\nhat'-r\nhat)}
\tilde{A}_{\rm RSD}^*(\mu,k\mu)
\tilde{A}_{\rm RSD}(\mu',k\mu')P_g(k)
\vs
=&
\sum_{np}\sum_{n'p'}
d_{np}d_{n'p'}
\int d\Omega_{\nhat} Y_{\ell m}(\nhat)
\int d\Omega_{\nhat'} Y_{\ell' m'}^*(\nhat')
\vs
&\times
\int\frac{d^3k}{(2\pi)^3} e^{i\vk\cdot(r'\nhat'-r\nhat)}
P_g(k) k^{n+n'}\mu^p{\mu'}^{p'}
\vs
=&
\sum_{np}\sum_{n'p'}
d_{np}d_{n'p'}
\int d\Omega_{\nhat} Y_{\ell m}(\nhat)
\int d\Omega_{\nhat'} Y_{\ell' m'}^*(\nhat')
\vs
&\times
\frac{\partial^{p'}}{\partial(ikr')^{p'}}
\frac{\partial^{p}}{\partial(-ikr)^{p}}
\int\frac{d^3k}{(2\pi)^3} e^{i(kr'\mu'-kr\mu)}
P_g(k) k^{n+n'}\,,
\ea
where we convert the $\mu$-dependences to derivatives.
We then use Rayleigh's formula
\be
e^{i\vk\cdot\vr}
=
4\pi\sum_{\ell m}i^\ell j_{\ell}(kr) Y_{\ell m}^*(\khat) Y_{\ell m}(\nhat)\,,
\label{eq:Rayleigh}
\ee
and the orthonormality of the spherical harmonics
\be
\int d\Omega_{\nhat}
Y_{\ell m}(\nhat)
Y_{\ell' m'}^*(\nhat) = \delta_{\ell\ell'}^K\delta_{mm'}^K\,,
\label{eq:YlmOrtho}
\ee
where $\delta^K_{ij}$ is the Kronecker delta. 
That simplifies the angular integrations and leads to the expression for the 
angular power spectrum
\ba
C_\ell(r,r')
=&\sum_{npn'p'}
i^{p-p'}d_{np}d_{n'p'}C_{\ell\ell}^{(p,p',n+n'+2)}(r,r')\,,
\label{eq:Clrr}
\ea
where we use $C_{\ell\ell'}^{(n,n',\alpha)}(r,r')$ defined earlier 
in \cref{eq:Cll}. The fundamental problem we encounter here is that,
for the galaxy power spectrum that scales as 
$\lim_{k\to\infty}P(k)\approxpropto k^{-\alpha}$,
the expression in \refeq{Clrr} does not converge for terms with 
$n+n'\gtrsim\alpha$. That is, for the linear galaxy power spectrum 
($\alpha=3$), the sum in \refeq{Clrr} diverges for all RSD terms with 
$n+n'\gtrsim3$.

This problem has not been addressed in literature thus far. 
Rather, in angular power spectrum analyses literature, 
the FoG are often ignored, since they mainly manifest themselves as a 
reduction in power on small scales
\cite{Castorina/White:2018,Tanidis+:2019arXiv190207226T}.
Others include the FoG as an additional redshift uncertainty
\cite{Loureiro+:2019MNRAS.485..326L}.

\section{Divergent-free expression for 
the angular power spectrum of galaxies}
\label{sec:wide-angle}
In this section, we resolve the problem by transforming the diverging integration appearing in \refeq{Clrr} to calculate the angular power spectrum including the nonlinear Kaiser effect.
To do so, let us introduce the radial window function $W_r(x)$ 
normalized as
\be
\int_0^\infty \dd x \, W_r(x) = 1\,,
\ee
with which we write the observed spherical harmonic coefficients as
\ba
\bdelta_{\ell m}^\mathrm{RSD}(r)
&= \int\dd r'
\, \delta^\mathrm{RSD}_{\ell m}(r')
\, W_r(r')
\vs
&= \int\dd r' \, W_r(r') \int\dd\Omega \, Y^*_{\ell m}(\nhat)
\int\frac{\dd^3k}{(2\pi)^3} \, e^{ikr'\mu}
\vs&\quad\times
\widetilde A_\mathrm{RSD}(\mu,k\mu) \, \delta_g^\mathrm{real}(\vk)\,.
\label{eq:delta_lm}
\ea
Here, $\widetilde{A}_\mathrm{RSD}(\mu,k\mu)$ is the RSD operator defined in \refeq{Arsd}.
Hereafter, we use $\bdelta_{\ell m}^{\rm RSD}$ to refer the 
harmonic coefficients  of the density field binned with the 
radial-window function.
For the sharp window function, $W_r(r')=\delta^D(r-r')$, we recover 
the expression for $\delta_{\ell m,g}^{\rm RSD}$ in \refeq{almr0}.

The key observation here is that we can make replacements,
$\mu\to-ik^{-1}\partial_{r'}$ and $k\mu\to-i\partial_{r'}$ 
both of which act on the exponential $e^{ikr'\mu}$, to re-write 
\refeq{delta_lm} as 
\ba
\label{eq:delta_g_deikr}
\bdelta_{\ell m}^\mathrm{RSD}(r)
   &=
\int\dd\Omega \, Y^*_{\ell m}(\rhat)
\int\frac{\dd^3k}{(2\pi)^3}
\, \delta^\mathrm{real}(\vk)
\vs&\times
\int\dd r' \, W_r(r')
\, \widetilde A_\mathrm{RSD}(-ik^{-1}\partial_{r'},-i\partial_{r'})
\, e^{ikr'\mu}\,.
\ea
We then use the integration by part
\citep{Assassi+:2017JCAP...11..054A,Schoneberg/etal:2018,DiDio+:2019JCAP...04..053D}
to move the derivative operator 
$\widetilde{A}_{\rm RSD}(-ik^{-1}\partial_{r'},-i\partial_{r'})$ acting on 
the exponential onto the window function.
That is, for each term in the series-expansion, 
\refeq{series-expansion}, performing the integration-by-parts $p$ times leads
to 
\ba
\label{eq:delta_g_dW(r)}
\bdelta_{\ell m}^\mathrm{RSD}(r)
&=
\int\dd\Omega \, Y^*_{\ell m}(\rhat)
\int\frac{\dd^3k}{(2\pi)^3}
\, \delta_g^\mathrm{real}(\vk)
\vs&\quad\times
\int\dd r' \, e^{ikr'\mu}
\, \widetilde A_\mathrm{RSD}(ik^{-1}\partial_{r'},i\partial_{r'})
\, W_r(r')\,.
\ea
The swap of the differential operator is valid as long as the window function
$W_r(r')$ vanishes at the boundaries ($r=0,\infty$), which is true for all
practical cases. Other than the constraints at the boundaries, we have the
freedom to choose the shape of the window function, or radial binning, for the
analysis.

Finally, using Rayleigh's formula [\refeq{Rayleigh}] and the orthogonality 
of the spherical harmonics [\refeq{YlmOrtho}], we find the expression for 
the angular power spectrum as 
\ba
   &\<\bdelta_{\ell m}^\mathrm{RSD}(r_1) \bdelta_{\ell'm'}^{\mathrm{RSD},*}(r_2)\>
\equiv
\delta^K_{\ell\ell'} \, \delta^K_{mm'} \, C_\ell^\mathrm{RSD}(r_1,r_2)
\vs
&=
\delta^K_{\ell\ell'} \, \delta^K_{mm'}
\int\dd r
\int\dd r'
\, \frac{2}{\pi} \int\dd k \, k^2 \, j_\ell(kr) \, j_{\ell'}(kr')
\, P_g(k)
\vs&\quad\times
\left[\widetilde A_\mathrm{RSD}(ik^{-1}\partial_r,i\partial_r) \, W_{r_1}(r)\right]
\vs&\quad\times
\left[\widetilde A_\mathrm{RSD}(-ik^{-1}\partial_{r'},-i\partial_{r'}) \, 
W_{r_2}(r')\right]\,.
\label{eq:Cl_RSD}
\ea
The Kronecker deltas signify the statistical homogeneity and isotropy.
It is obvious that the troublesome divergent integrals in \refeq{Clrr}
disappear in \refeq{Cl_RSD} for a sufficiently differentiable 
radial window function. Instead, \refeq{Cl_RSD} shows that the
effect of RSDs can be captured by a $k$-dependent change of the window
function. That is, in spherical harmonic space, RSD distorts the shape of 
the redshift binning, or radial window function; we can model the RSD effect 
by taking into account the distortion of the window function. 

From the fact that the RSD effect comes as the derivative operator acting on the
window function, we can already deduce some useful facts.
If the window function is sharply peaked, then the derivatives will be large,
and the RSD effect should be large. Conversely, a broader window function 
would yield a smaller RSD effect.

As each $k$ from a higher-order term adds a derivative on the 
window function,
\cref{eq:Cl_RSD} only converges if the window function is sufficiently smooth
(differentiable) so that the high-$k$ limit is suppressed. Indeed, 
as we show later in \cref{eq:Wtilde} for the flat-sky approximation,
even a top-hat window function leads to some suppression of high-$k$ modes. 
For the non-linear RSD that we consider here, the FoG introduce a natural 
high-$k$ cutoff so that \cref{eq:Cl_RSD} is finite even for a 
Dirac-delta window function. For more general cases, a phenomenological 
expansion containing progressively higher powers in $k$ due to 
high-order terms can be compensated by choosing a sufficiently 
smooth window function.

Considering the derivation in \refeqs{delta_g_deikr}{delta_g_dW(r)}, note that
it is by no means necessary to move all derivatives in $\widetilde
A_\mathrm{RSD}(ik^{-1}\partial_r,i\partial_r)$ onto the window function. For
example, we may choose to leave the operators related to the linear Kaiser
effect $\widetilde{A}_{\rm RSD}^{\rm linear}= 1 - \beta k^{-2}\partial_{r}^2$
[see \cref{eq:Arsd}]
as a derivative on the Fourier kernel $e^{ikr\mu}$:
\ba
\label{eq:delta_g_dW(r)_LNL}
&\bdelta_{\ell m}^\mathrm{RSD}(r)
=
\int\dd\Omega \, Y^*_{\ell m}(\rhat)
\int\frac{\dd^3k}{(2\pi)^3}
\, \delta_g^\mathrm{real}(\vk)
\vs&\times
    \int\dd r' \,
    \bigg[\widetilde A_\mathrm{nl}(ik^{-1}\partial_{r},i\partial_{r'}) W_r(r')\bigg]
    \bigg[(1-\beta k^{-2}\partial_{r'}^2)e^{i\vk\cdot\vr'}\bigg]
\,.
\ea
In fact, we find that separating the linear and nonlinear RSD effects as in 
\refeq{delta_g_dW(r)_LNL} eases the numerical implementation, and 
simplifies the subsequent analysis. In practice, that also allows us to 
treat the FoG and other nonlinear corrections as a modification to the window 
function entirely separate from the linear Kaiser effect. 

\refeqs{Cl_RSD}{delta_g_dW(r)_LNL} are the main results of this paper. 
In the rest of the paper, we shall present the result of numerical 
implementation of these equations. Comparing the result with the small-angular
scale correlation function, we also find a simple interpretation of the 
harmonic-space galaxy power spectrum in terms of the usual Fourier-space 
power spectrum.

One note on the implementation of \refeq{Cl_RSD} is in order here.
Often in literature to calculate the linear RSD effect, the integrations 
over the window functions in \cref{eq:Cl_RSD} are pulled under the $k$-intergral. 
That would have the advantage that the integration over the window function
needs to be performed only once. 
In the new formalism for the nonlinear Kaiser effect, this is not true anymore with the $k$-dependent modification of the window function.
Moreover, in that way, the $k$-integral 
still requires integration over a highly oscillatory function and it 
precludes the use of the 2-FAST algorithm. 
To take full advantage of the 2-FAST algorithm, we shall execute the 
$k$-integral first, then apply the 
window functions afterwards.

\subsection{Convolved window function}
\label{sec:winfn}
\begin{figure*}
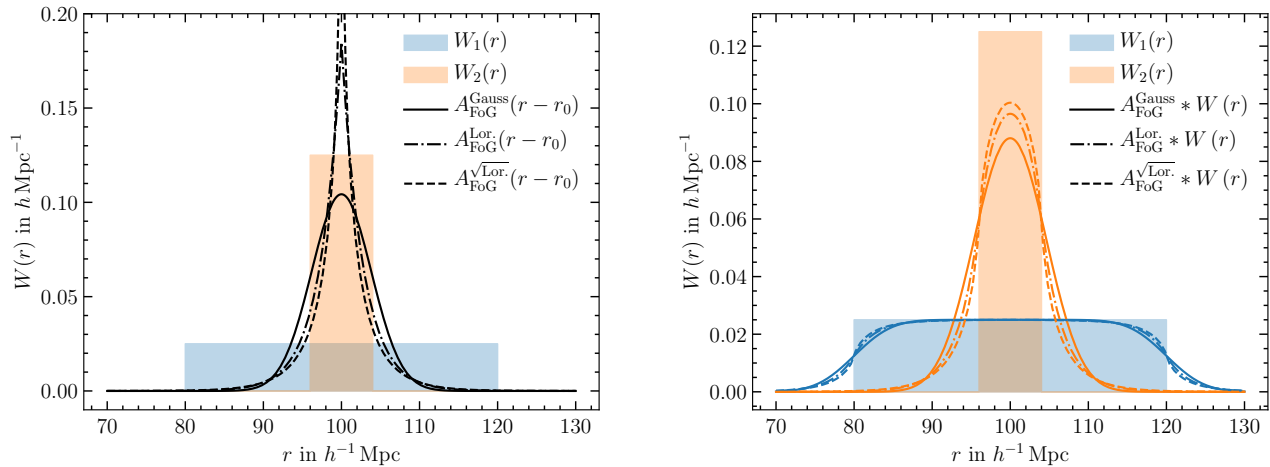

    \centering
    \incgraph{{figs/winfn_Afog}.pdf}
    \incgraph{{figs/winfn_eff_gaussian_lorentzian}.pdf}
    \caption{Left: Illustration of a wide top-hat window function $W_1(r)$
        (blue shaded area), a narrow top-hat ($W_2(r)$, orange shaded area),
        and the three convolution kernels \refeqs{Afogr_a}{Afogr_c}
        corresponding to the three models \refeqs{Afog_a}{Afog_c}.
        Right: The shaded areas are the same as in the left plot. If the window
        function in redshift space is a top-hat, then the real-space window
        function is a smoothed top-hat given by the convolution
        $A_\mathrm{FoG}*W\,(r)$, shown here for the same window functions and
        FoG models as in the left plot, see \refeqs{AfogW_a}{AfogW_c}.
    }
    \label{fig:winfn_eff}
\end{figure*}
The nonlinear RSD kernels $\widetilde{A}_{\rm nl}(\mu,k\mu)$ in \refeqs{Afog_a}{Afog_c} only depend on $k\mu$, which yields one further simplication when computing 
the modified window function $\widetilde{A}_{\rm nl}(i\partial_{r'})W_r(r')$.
Expressing the window function in terms of its Fourier transform
$\widetilde{W}(q)$, we find that the modified window function is given as a convolution
\ba
\widetilde A_\mathrm{nl}(i\partial_{r'}) \, W_r(r')
&= \int_{-\infty}^\infty\frac{\dd q}{2\pi}
\, \widetilde A_\mathrm{nl}(-q)
\, \widetilde{W}_r(q)
\, e^{iqr'}
\vs
&=
\int_{-\infty}^\infty\dd y
\, A_\mathrm{nl}(r'-y)
\, W_r(y)
\vs
&\equiv A_\mathrm{nl}(r') * W_r \, (r')
\,.
\label{eq:AfogOp_Winfn}
\ea
In deriving \refeq{AfogOp_Winfn} we assumed that the domain of the window
function, which is strictly speaking only defined for $r\geq0$, can be
extended to negative $r$ as well, and that it vanishes there.

Here, $A_\mathrm{nl}(r)$ is the inverse Fourier transform of $\widetilde
A_\mathrm{nl}(q)$. \refeq{AfogOp_Winfn} shows that the effective real-space window function is the
radial convolution of the window function with the nonlinear RSD operator. 
The meaning of the RSD modification of the window function may be most apparent
when applying the Fingers-of-God operators in \refeqs{Afog_a}{Afog_c}, for which 
the corresponding real-space functions are given by 
\ba
\label{eq:Afogr_a}
A_\mathrm{FoG}^{\rm Gauss}(r'-r)
&= \frac{1}{\sqrt{2\pi}\, \sigma_u} \, e^{-\frac{(r'-r)^2}{2 \sigma_u^2}}\,,
\\
\label{eq:Afogr_b}
A_\mathrm{FoG}^{\rm Lor.}(r'-r)
&= \frac{1}{\sqrt2\,\sigma_u}\,e^{-\frac{\sqrt2}{\sigma_u}\,|r'-r|}\,,
\\
\label{eq:Afogr_c}
A_\mathrm{FoG}^{\rm \sqrt{Lor.}}(r'-r)
&= \frac{1}{\pi \sigma_u} \, K_0\!\(\frac{|r'-r|}{\sigma_u}\)\,,
\ea
where $K_n(x)$ is a modified Bessel function of order $n$
\citep{Taylor+:2001MNRAS.327..689T}.
Note that \refeqs{Afogr_a}{Afogr_c} are simply the assumed 1D radial 
velocity distibution for each FoG model. 
The modified window function, therefore, incorporates the galaxies moving 
from the adjacent bins with the probability given by the velocity 
dispersion function 
\citep{Kang+:2002MNRAS.336..892K,Percival/etal:2004,Loureiro+:2019MNRAS.485..326L}.

For definiteness, we consider a case of radial binning with a top-hat window
function of width $\Delta r_i$ centered on $r_i$: $W_i(r)=\frac{1}{\Delta
r_i}$ when $r_i^\mathrm{lo}\leq r\leq r_i^\mathrm{hi}$ and vanishes otherwise,
where $r_i^\mathrm{lo}=r_i-\frac12\Delta r_i$ and
$r_i^\mathrm{hi}=r_i+\frac12\Delta r_i$ are the lower and upper bounds of the
bin $i$. Then the modifed window functions given by the convolution in
\refeq{AfogOp_Winfn} are
\ba
\label{eq:AfogW_a}
A^{\rm Gauss}_\mathrm{FoG} * W_i \, (r)
&= \frac{\erf\!\(\frac{r_i^\mathrm{hi}-r}{\sqrt2\, \sigma_u}\)}{2\Delta r_i}
- \frac{\erf\!\(\frac{r_i^\mathrm{lo}-r}{\sqrt2\, \sigma_u}\)}{2\Delta r_i}\,,
\\
\label{eq:AfogW_b}
A^{\rm Lor.}_\mathrm{FoG}*W_i\,(r)
&=
\frac{1}{2\Delta r_i}
\, \frac{r_i^\mathrm{hi}-r}{|r_i^\mathrm{hi}-r|}\(1 - e^{-\frac{\sqrt2}{\sigma_u}|r_i^\mathrm{hi}-r|}\)
\vs&\quad
- \frac{1}{2\Delta r_i}
\, \frac{r_i^\mathrm{lo}-r}{|r_i^\mathrm{lo}-r|}\(1 - e^{-\frac{\sqrt2}{\sigma_u}|r_i^\mathrm{lo}-r|}\) \,,
\\
\label{eq:AfogW_c}
A^{\rm \sqrt{Lor.}}_\mathrm{FoG} * W_i\,(r)
&=
\frac{1}{\pi \, \Delta r_i}
\int^{\frac{r_i^\mathrm{hi}-r}{\sigma_u}}_{\frac{r_i^\mathrm{lo}-r}{\sigma_u}}
\dd x \, K_0(|x|)\,,
\ea
where $\erf(x)\equiv\frac{2}{\sqrt\pi}\int_0^x \dd t\, e^{-t^2}$ is the error
function. The integration of the modified Bessel function in
\refeq{AfogW_c} can be expressed using the following identity:
\ba
\frac{1}{\pi} \int_0^x \dd x' \, K_0(x')
 &= \frac12\,x\, K_0\!\(x\) \, \mathbf{L}_{-1}\!\(x\)
\vs&\quad
+ \frac12\,x\, K_1\!\(x\) \, \mathbf{L}_{0}\!\(x\)\,,
\ea
where ${\bf L}_n(x)$ is the modified Struve function of order $n$.

Should one use the phenomenological nonlinear RSD terms such as 
$\sum_{a,b}k^a\mu^b$ multiplied with the FoG factors, 
for example, as done in \cite{shoji/etal:2009}, one may need to calculate
higher-order derivatives of the convolved window functions in \refeqs{AfogW_a}{AfogW_c}. 

The left panel of \reffig{winfn_eff} illustrates the convolution kernels
\refeqs{Afogr_a}{Afogr_c} for ($r_0=100\,~{\rm Mpc}/h$) and $\sigma_u=\SI{3.8}{\per\h\mega\parsec}$. 
For comparison we also show a wide top-hat bin of width $\Delta r=\SI{40}{\per\h\mega\parsec}$ (blue shaded box) and a narrow bin with 
$\Delta r=\SI{8}{\per\h\mega\parsec}$ (orange shaded box).
We also show the modified window functions for the three FoG models and for the 
same two example top-hat functions in the right panel of \reffig{winfn_eff}.

The edges of the top-hat window function are smoothed by the convolution. 
This means that the galaxies contained in the top-hat bin $W(r)$ defined in 
the redshift-space are selected with a probability proportional 
to $A_\mathrm{RSD} * W\,(r)$ in real space. 
As expected, the modification of the window function 
(thus, the nonlinear RSD effect) is bigger for narrower window functions. 
For the narrow-window-function example ($\Delta r=8\,h^{-1}{\rm Mpc}$), about one 
third of the galaxies come from outside the top-hat boundaries. 
For the wide example ($\Delta r = 40\,h^{-1}{\rm Mpc}$), only the edges 
are changed, so that only \SI{\sim8}{\percent} of galaxies are 
different between real space and redshift space. These effects 
are largest for Gaussian FoG, and smallest for square-root Lorentzian FoG.

\section{Result: \texorpdfstring{$C_\ell$}{C\_ell} and \texorpdfstring{$P(k_\perp)$}{P(k)}}
\label{sec:flatsky}
\begin{figure*}
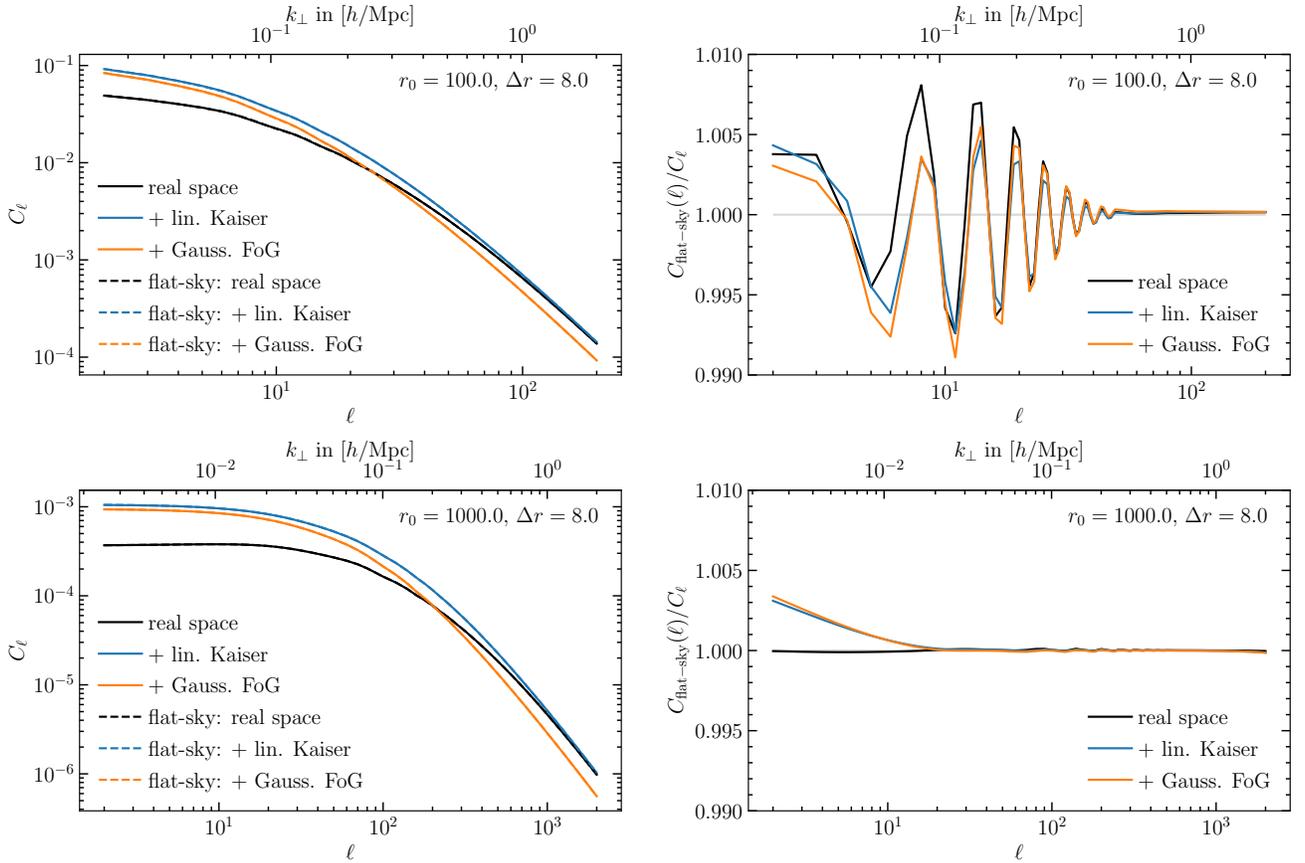

    \centering
    \incgraph[0.49]{{figs/cell_tophat_r100.0_s4.0_th_ln_fog_r100_D1.0_b1.0_f0.541_sv25.0}.pdf}
    \incgraph[0.49]{{figs/cell_ckperp_ratio_tophat_r100.0_s4.0_th_ln_fog_r100_D1.0_b1.0_f0.541_sv25.0}.pdf}
    \incgraph[0.49]{{figs/cell_tophat_r1000.0_s4.0_th_ln_fog_r1000_D1.0_b1.0_f0.706_sv25.0}.pdf}
    \incgraph[0.49]{{figs/cell_ckperp_ratio_tophat_r1000.0_s4.0_th_ln_fog_r1000_D1.0_b1.0_f0.706_sv25.0}.pdf}
    \caption{
        Comparison between the harmonic-space power spectra and the flat-sky
        approximation. The top two panels are for a top-hat window function 
        of width
        $\Delta r=\SI{8}{\per\h\mega\parsec}$ centered around
        $r_0=\SI{100}{\per\h\mega\parsec}$, the bottom two panels are for
        $r_0=\SI{1000}{\per\h\mega\parsec}$ with the same width.
        Left panels: solid lines show the exact calculation, the dashed
        lines assume the flat-sky approximation.
        Right panels: the ratio between the flat-sky approximation and the
        exact calculation for the same narrow window function. To achieve
        below-percent-level agreement we use the correspondence
        \refeq{Cellkperp}. This same level of agreement is achieved with the
        Lorentzian and square-root-Lorentzian FoG models.
        All plots show the corresponding flat-sky $k_\perp$ mode on
        top.
    }
    \label{fig:compare_fullsky_flatsky}
\end{figure*}
\begin{figure*}
    \centering
    \incgraph[0.49]{{figs/cell_tophat_r100.0_s20.0_th_ln_fog_r100_D1.0_b1.0_f0.541_sv25.0}.pdf}
    \incgraph[0.49]{{figs/cell_ckperp_ratio_tophat_r100.0_s20.0_th_ln_fog_r100_D1.0_b1.0_f0.541_sv25.0}.pdf}
    \incgraph[0.49]{{figs/cell_tophat_r1000.0_s20.0_th_ln_fog_r1000_D1.0_b1.0_f0.706_sv25.0}.pdf}
    \incgraph[0.49]{{figs/cell_ckperp_ratio_tophat_r1000.0_s20.0_th_ln_fog_r1000_D1.0_b1.0_f0.706_sv25.0}.pdf}
    \caption{
        Same as \cref{fig:compare_fullsky_flatsky} but with a bin width
        $\Delta r=\SI{40}{\per\h\mega\parsec}$.
    }
    \label{fig:compare_fullsky_flatsky_sigma20}
\end{figure*}
\begin{figure*}
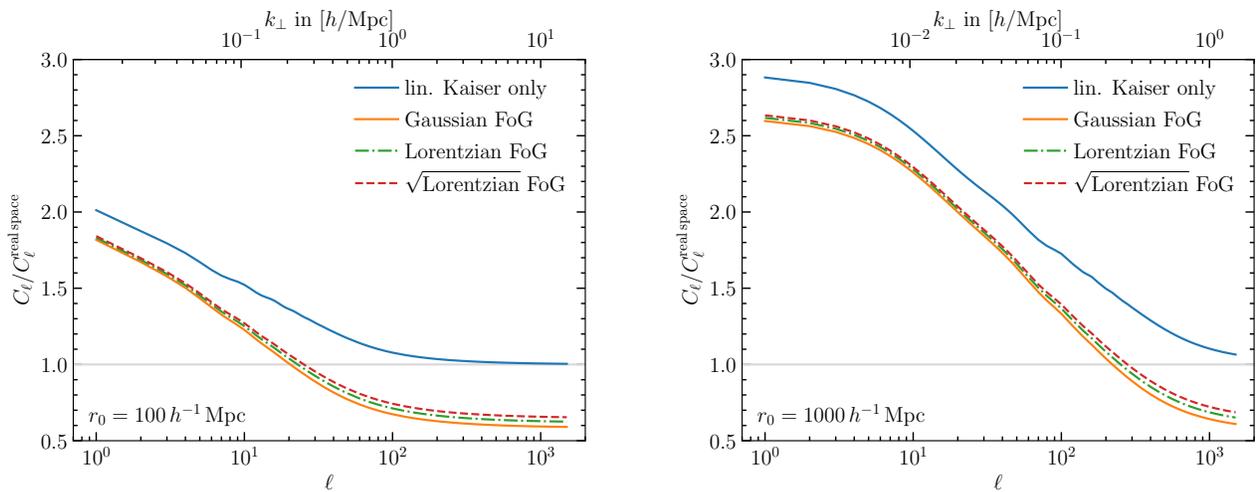

    \centering
    \incgraph{{figs/cl_compare_ratios_r100.0_dr8.0}.pdf}
    \incgraph{{figs/cl_compare_ratios_r1000.0_dr8.0}.pdf}
    \caption{Ratio of RSD angular power spectra to the real-space angular
        power spectrum, for $r_0=\SI{100}{\per\h\mega\parsec}$ (left) and
        $r_0=\SI{1000}{\per\h\mega\parsec}$ (right). The bin width for both
        plots is $\Delta r=\SI{8}{\per\h\mega\parsec}$. For the three FoG
        models, the velocity dispersion is the same, and it is chosen according
        to \refeq{sigma_u}. The limited width of the redshift bin introduces
        non-linearities such as the FoG on all transverse scales $\ell$.
        The corresponding transverse $k_\perp$-mode is shown at the
        top.
    }
    \label{fig:compare_ratios}
\end{figure*}
With the modified window functions shown in \reffig{winfn_eff},
we now compute the shape of the harmonic-space power spectrum
with nonlinear redshift-space distortion.
To apply the 2-FAST algorithm \citep{GrasshornGebhardt+:2018PhRvD..97b3504G},
we transform the integral over $r'$ in \refeq{Cl_RSD} to an integral over the
ratio $R=r'/r$.

Along with the full harmonic space expression, we also compute the 
power spectrum with the flat-sky approximation. In the flat-sky calculation,
we keep constant $\hat{z}$-direction throughout the volume, and compute 
the harmonic space powerspectrum by projecting the three-dimensional 
power spectrum along the parallel (line-of-sight) direction.
The implementation of flat-sky approximation is easier as two of the 
three integrals in \refeq{Cl_RSD} can be done analytically.
As we show in the following section, the flat-sky approximation provides 
a good approximation when matching $\ell= k_\perp r_0 - \frac12$ between the
multipole moment and three-dimensional transverse Fourier wavenumber.

\subsection{Fourier-space expression with the flat-sky approximation}
With the flat-sky approximation, we obtain the tangential
two-dimensional ($\vx_\perp$) density contrast by integrating the 
three-dimensional density contrast along the line-of-sight,
\ba
\delta_s(\vx_\perp) &= \int \dd z \, W(z) \, \delta_s(\vx)\,,
\ea
where $\delta_s(\vx)$ is the redshift space density contrast, and $W(z)$ is 
the radial window function.
The Fourier-space density contrast is then,
\ba
\delta_s(\vk_\perp)
&= \int \dd^2x_\perp \, \delta_s(\vx_\perp) \, e^{-i\vk_\perp\cdot\vx_\perp}\,.
\ea
Expressing the density contrast $\delta_s(\vx)$ in
terms of its Fourier components $\delta_s(\vk)$ allows us to perform the
integrals over ${\vx}_\perp$ analytically. We get
\ba
\delta_s(\vk_\perp)
&= 
\int \dd^2x_\perp
\left[\int dz \, W(z)\int \frac{d^3q}{(2\pi)^3} 
\, \delta_s(\vq) \, e^{i\vq\cdot\vx}
\right] 
e^{-i\vk_\perp\cdot\vx_\perp}
\vs
&=
\int\frac{\dd k_z}{2\pi}
\, \delta(\vk)
\, \widetilde A_\mathrm{RSD}\(\frac{k_z}{k},k_z\)
\, \widetilde{W}^*(k_z)\,,
\label{eq:delta_s(k_perp)}
\ea
where we used \refeq{delta_g}, $\mu\equiv\zhat\cdot\khat$, and $\widetilde
W(k_z)$ is the Fourier transform of the window function. 
Defining the perpendicular two-dimensional power spectrum as
\ba
\<\delta_s(\vk_\perp) \delta_s^*(\vk'_\perp)\>
&= (2\pi)^2 \delta^D(\vk_\perp-\vk'_\perp) \, C(k_\perp)\,,
\ea
we find that 
\ba
C(k_\perp;r_1,r_2)
&=
\int\frac{\dd k_z}{2\pi}
\, P(k)
\, \widetilde{W}^*_1(k_z)
\, \widetilde{W}_2(k_z)
\vs&\quad\times
\widetilde A_\mathrm{RSD}^{r_1}\!\(\frac{k_z}{k},k_z\)
\, \widetilde A_\mathrm{RSD}^{r_2*}\!\(\frac{k_z}{k},k_z\)\,,
\label{eq:flat_sky}
\ea
with $k=\sqrt{k_\perp^2 + k_z^2}$, and the superscript $r_i$ in $\widetilde
A_\mathrm{RSD}$ indicates the radial-dependence of the coefficients, for
example $f(r_i)$ and $\sigma_u(r_i)$, of $\widetilde{A}_{\rm RSD}$.
Note that, in \cref{eq:flat_sky} we assume that the power spectrum $P(k)$ 
does not depend on redshift, but we can easily include the time-dependence into
the $\widetilde{A}_{\rm RSD}$.
For example, the linear growth factor $D(r_1)D(r_2)$ would introduce
a constant multiplication factor to $\widetilde{A}_{\rm RSD}$.

In order to relate \refeq{flat_sky} to the angular power spectrum, we convert
the two-dimensional Fourier wavenumber $k_\perp$ to the harmonic space moment
$\ell$ as, $\ell + \frac12=k_\perp r_0$,
\citep{GrasshornGebhardt+:2018PhRvD..97b3504G} 
\footnote{In short, it is motivated by matching the eigenvalues of the angular Laplacian $\nabla^2_\theta$ and the two-dimensional Laplacian $\nabla^2_\perp$:
$r_0^2 k_\perp^2 = \ell(\ell+1)$:
$k_\perp r_0 = \sqrt{\ell(\ell+1)} = \ell(1+1/\ell)^{1/2} \simeq \ell+1/2 + 
\mathcal{O}(1/\ell)$.
}, where
$r_0\equiv\frac12(r_1+r_2)$, and 
\ba
C_\ell
   &= \frac{1}{r_1 r_2} \, C\(k_\perp=\frac{\ell+1/2}{r_0}\)\,.
   \label{eq:Cellkperp}
\ea
For a top-hat window function of width $\Delta r_i$ centered around $r_i$, 
we have $W_i(z)=1/\Delta r_i$, and the Fourier transform is
\ba
\label{eq:Wtilde}
\widetilde W_i(k_z)
&=
\, e^{-ik_z r_i}
\, j_0\!\(\frac{k_z\Delta r_i}{2}\)\,,
\ea
where $j_0(x)\equiv \sin(x)/x$ is the spherical Bessel function of 
order 0. Therefore, the cross-correlation between two bins of widths $\Delta r_1$ and $\Delta r_2$
centered on $r_1$ and $r_2$ is in the flat-sky approximation given by
\ba
C(k_\perp, r_1, r_2)
&=
\int\frac{\dd k_z}{2\pi}
\, P(k)
\, \cos\!\(k_z (r_1-r_2)\)
\vs&\quad\times
j_0\!\(\frac{k_z\Delta r_1}{2}\)
\, j_0\!\(\frac{k_z\Delta r_2}{2}\)
\vs&\quad\times
\widetilde A_\mathrm{RSD}^{r_1}\!\(\frac{k_z}{k},k_z\)
\widetilde A_\mathrm{RSD}^{r_2*}\!\(\frac{k_z}{k},k_z\)
\,,
\label{eq:Clflat_kperp}
\ea
where the imaginary part vanishes since all terms other than the exponential
are even in $k_z$, and we assume that the RSD factor is real, e.g., as in
\cref{eq:Arsd,eq:Afog_a,eq:Afog_b,eq:Afog_c}.
Using \refeq{Clflat_kperp}, we find the auto-correlation function as
\ba
C(k_\perp, r_0)
&=
\int\frac{\dd k_z}{2\pi}
\, P(k)
\left[
    j_0\!\(\frac{k_z\Delta r}{2}\)
\widetilde A_\mathrm{RSD}^{r_0}\!\(\frac{k_z}{k},k_z\)
\right]^2
\,,
\label{eq:Clflat_kperp_auto}
\ea
where we set $r_0=r_1=r_2$ and $\Delta r=\Delta r_1=\Delta r_2$.

\subsection{Small-scale (\texorpdfstring{$k_\perp\to\infty$}{k\_perp -> infty} or $\ell\to\infty$) limit}
\label{sec:smallscale}
In the small-tangential (angular) scale limit where 
$k_\perp\to\infty$, we get for the auto-correlation
\ba
\label{eq:kperp_to_infty_flat_sky}
\lim_{k_\perp\to\infty} C(k_\perp)
&=
P(k_\perp)
\int\frac{\dd k_z}{2\pi}
\, \left|\widetilde A_\mathrm{RSD}\!\(0,k_z\)
\, \widetilde{W}(k_z) \right|^2\,.
\ea
That is, the suppression of the power spectrum due to FoG becomes independent
of $k_\perp$, or $\ell$. As the flat-sky approximation is valid on small 
scales, we expect that the same is true for the exact calculation as well.
The suppression factor for a top-hat window function and
Gaussian FoG relative to real space only depends on the width of the window function $\Delta r$ and the velocify dispersion $\sigma_u$:
\ba
&\frac{
\int\frac{\dd k_z}{2\pi}
\, \left|\widetilde A_\mathrm{RSD}\!\(0,k_z\)
\, \widetilde{W}(k_z) \right|^2
}{
\int\frac{\dd k_z}{2\pi}
\, \left|\widetilde{W}(k_z) \right|^2
}
\vs
&=
\erf\!\(\frac{\Delta r}{2\sigma_u}\)
- \frac{2\sigma_u}{\sqrt{\pi}\,\Delta r}
\(1 - e^{-\frac{\Delta r^2}{4 \sigma_u^2}}\)
\,.
\ea
Similar expressions can be found for other forms of the FoG.

\subsection{Nonlinear RSD in Harmonic space \texorpdfstring{$C_\ell$}{C\_ell}}
In \reffig{compare_fullsky_flatsky} we show the harmonic-space 
power spectra calculation for a window function of width $\Delta
r=\SI{8}{\per\h\mega\parsec}$ centered around
$r_0=\SI{100}{\per\h\mega\parsec}$ (top panels) 
and we repeat this for a window function of the same width centered 
around $r_0=\SI{1000}{\per\h\mega\parsec}$ (bottom panels). 
For each case, we show the real-space power spectrum, the RSD power 
spectrum with only the linear Kaiser effect (without 
$\widetilde{A}_{\rm nl}$ in \refeq{Arsd}), and the power spectrum
that includes the linear Kaiser effect and Gaussian FoG.

In \reffig{compare_fullsky_flatsky}, we notice a few RSD features in 
harmonic space with narrow radial binning. 
First, as we expect from the three-dimensional RSD, the linear Kaiser 
effect enhances the power spectrum on large scales. 
The linear Kaiser effect, however, in harmonic space shows a strong
scale-dependence, and the enhancement vanishes on small scales.
Second, unlike the three-dimensional RSD, the Fingers-of-God effect reduces the 
power spectrum on {\it all} scales, but more so on small scales.
This is because the modified window function affects the angular clustering 
on all scales.

In addition, \reffig{compare_fullsky_flatsky} shows that the flat-sky
approximation (dashed line) agrees quite well with the exact result in 
harmonic space (solid line) on all scales. 
As shown in the right panels of \reffig{compare_fullsky_flatsky},
\refeq{Clflat_kperp_auto} leads to an agreement between the full formula 
and the flat-sky formula better than \SI{0.8}{\percent} for the narrow 
window function considered here.
The bottom panel shows that the flat-sky approximation proves to be more accurate at the larger radius $r_0=\SI{1000}{\per\h\mega\parsec}$.
With a wider window function $\Delta r=\SI{40}{\per\h\mega\parsec}$ as shown in \cref{fig:compare_fullsky_flatsky_sigma20} the differences become larger.
We also find that the agreements between the exact and 
flat-sky calculations holds the same for the Lorentzian and
square-root-Lorentzian FoG cases.
Note the sub-percent deviation at high $\ell$ for the $\Delta r=40\,h^{-1}{\rm Mpc}$ case 
shown in the top-right panel of \cref{fig:compare_fullsky_flatsky_sigma20}.
As the analysis in \cref{sec:limber} below shows, the discrepancy comes from the large $\Delta r/r$ 
for which the flat-sky approximation breaks. 
Nevertheless, the difference stays quite small even for this rather 
pathological example with $r=100\,h^{-1}{\rm Mpc}$ and $\Delta r =40\,h^{-1}{\rm Mpc}$.

Given the excellent agreement between the exact calculation and the flat-sky
approximation, we can understand the FoG effect on large angular scales 
as follows. In the $k_\perp\to0$ limit, the flat-sky formula gives 
(for Gaussian FoG as an example here)
\ba
C(0,r_0) &= (1+\beta)^2\int\frac{\dd k_z}{2\pi} \, P(k_z)
\, j_0^2\!\(\frac{k_z \Delta r}{2}\) e^{-\sigma_u^2k_z^2}
\ea
The spherical Bessel ensures that all modes up to $k_z\lesssim\frac{\pi}{\Delta
r}$ contribute, while the FoG suppression factor, on the other hand, affects
scales $k_z\gtrsim 1/\sigma_u$. 
The large-angular scale power spectrum is affected by the FoG effect 
if $1/\sigma_u \lesssim \pi/\Delta r$, or $\Delta r \lesssim \pi \sigma_u$.
For example, when $\sigma_u\sim3{\rm Mpc}/h$, 
the large angular-scale power spectrum for  
$\Delta r=8\,{\rm Mpc}/h < \pi \sigma_u = 10 {\rm Mpc}/h$ 
must be affected by FoG, but not for 
$\Delta r=40\,{\rm Mpc}/h > \pi \sigma_u = 10 {\rm Mpc}/h$.
That is consistent with what we observe in 
\reffigs{compare_fullsky_flatsky}{compare_fullsky_flatsky_sigma20}.

In \reffig{compare_ratios}, we compare the three forms for the FoG by
showing the ratio of the RSD angular power spectrum to the real-space angular
power spectrum in each case. Additionally, the figure shows the ratio for the
Kaiser effect only, and in the left panel we use
$r_0=\SI{100}{\per\h\mega\parsec}$ and in the right panel
$r_0=\SI{1000}{\per\h\mega\parsec}$.

Again, \reffig{compare_ratios} shows that the Kaiser effect vanishes on 
small scales, and the FoG, while present on all scales, is strongest on 
small scales.
Furthermore, the three forms of the FoG are very similar. As may be expected
from \reffig{winfn_eff}, Gaussian FoG are strongest while a square-root
Lorentzian is weakest for the same $\sigma_u$. The functional form is also
different in that a Gaussian FoG has a larger difference between large and
small scales than the other two. We have checked that this also holds true
even if $\sigma_u$ is adjusted so that the three forms agree 
on small scales using the analytical formula in \refsec{smallscale}.

\subsection{Limber's approximation}
\label{sec:limber}
\begin{figure*}
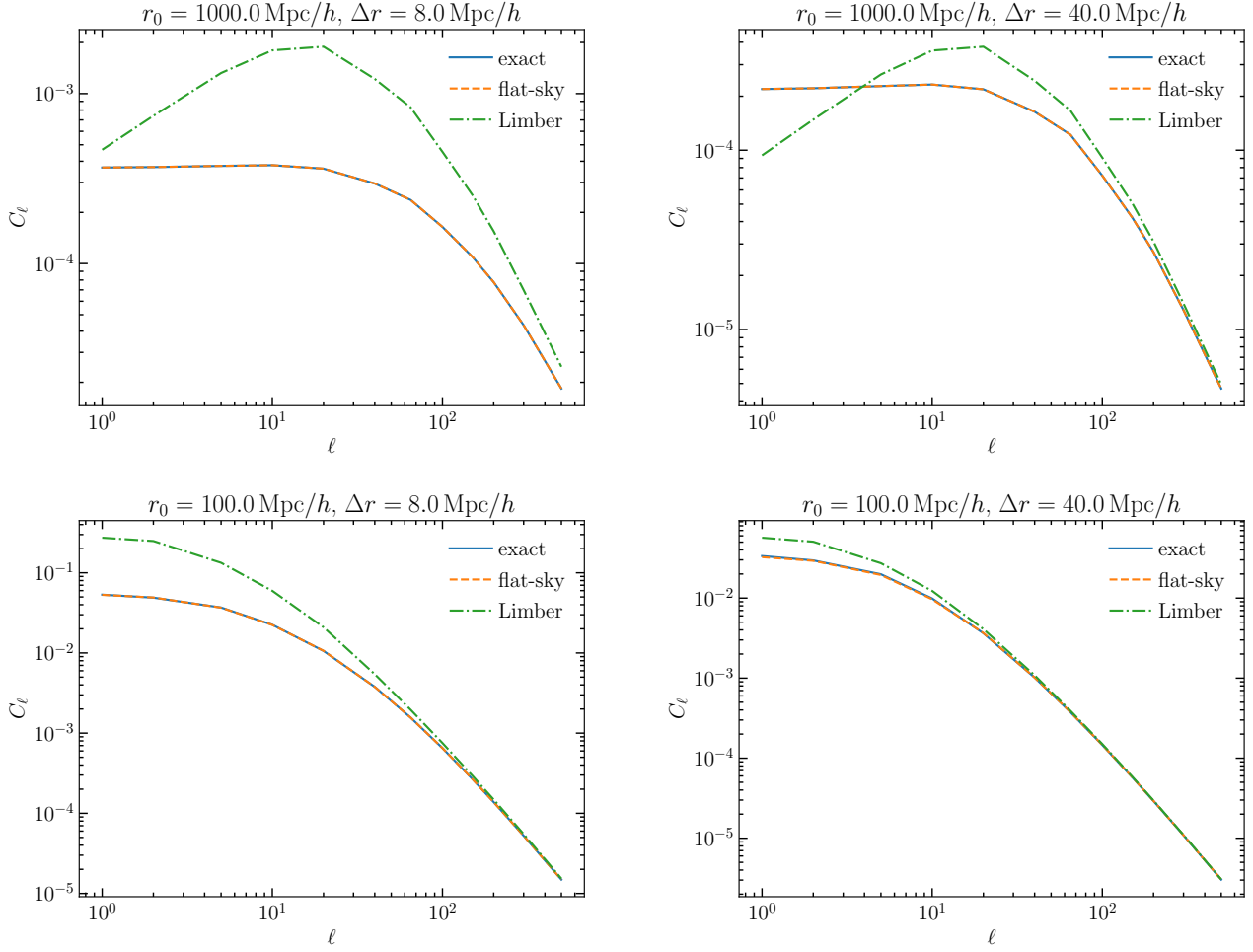

    \centering
    \incgraph{{figs/cl_exact_flat_limber_r01000.0_dr8.0}.pdf}
    \incgraph{{figs/cl_exact_flat_limber_r01000.0_dr40.0}.pdf}
    \incgraph{{figs/cl_exact_flat_limber_r0100.0_dr8.0}.pdf}
    \incgraph{{figs/cl_exact_flat_limber_r0100.0_dr40.0}.pdf}
    \caption{
        The real-space $C_\ell$ calculated using three different formulas: In
        blue the exact formula, in orange the flat-sky (almost directly on top
        of the blue line), in green Limber's approximation. Limber's
        approximation works better when the bin width is large.
        The corresponding $k_\perp$-mode is obtained by
        $k_\perp=(\ell-\frac12)/r_0$.
    }
    \label{fig:cl_exact_flat_limber}
\end{figure*}
\begin{figure*}
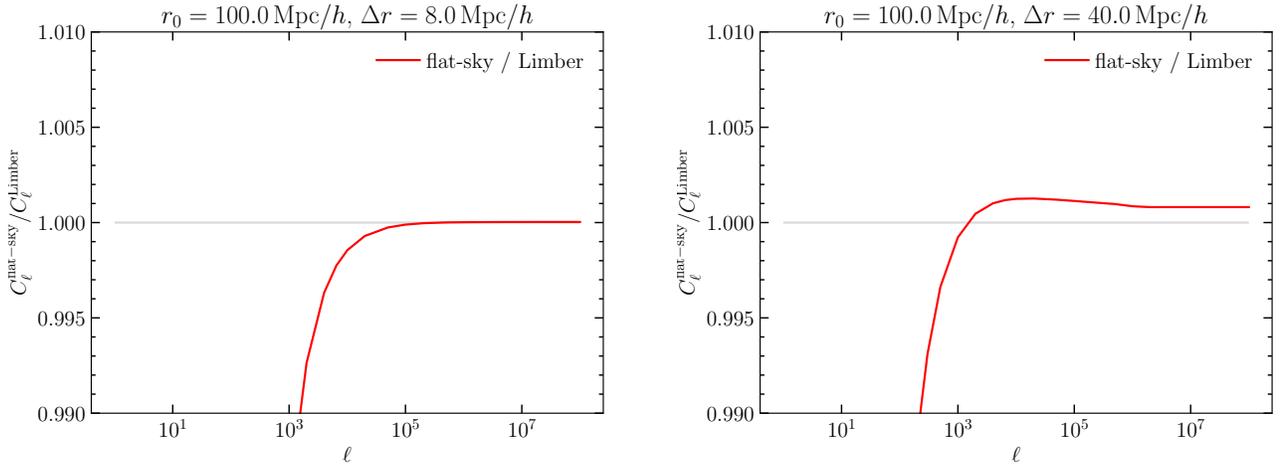

    \centering
    \incgraph{{figs/cl_flat_limber_r0100.0_dr8.0_ratio}.pdf}
    \incgraph{{figs/cl_flat_limber_r0100.0_dr40.0_ratio}.pdf}
    \caption{Ratio of flat-sky to Limber in real space. Only at very high
        $\ell$ is the predicted difference between flat-sky and Limber
        \cref{eq:limber_limit} reached. This figure assumes that the power
        spectrum remains a power-law far into the non-linear regime.
    }
    \label{fig:cl_flat_limber_ratio}
\end{figure*}
The top-right panel of \cref{fig:compare_fullsky_flatsky_sigma20} shows a 
constant discrepancy between the full calculation and the flat-sky 
approximation. In this section, we study the origin of this difference by
comparing the flat-sky approximation and the Limber approximation which 
provides an accurate approximation for large $\ell$.

Limber's approximation may be written as \citep{Loverde+:2008PhRvD..78l3506L}
\ba
j_\ell(kr) \to \sqrt{\frac{\pi}{2kr}}\,\delta^D\!\(kr - \ell-\frac12\)\,.
\ea
Then, \cref{eq:Cl_RSD} in real space for an auto-correlation can be approximated as
\ba
C_\ell^\mathrm{Real\,space}
&=
\int\dd r \, \frac{W^2\!\(r\)}{r^2}
\, P_g\!\(\frac{\ell+\frac12}{r}\),
\ea
and narrow window functions will enforce that 
$k\simeq\frac{1}{r_0}(\ell+\frac12)$, where $r_0$ is the radius to the bin center. 
For a power-law power spectrum $P(k)\propto k^{-(3+\epsilon)}$ and top-hat
window we then get, to first order:
\ba
C_\ell^\mathrm{Real\,space}
&=
\frac{1}{r_0^2\Delta r} \,P_g\!\(\frac{\ell+\frac12}{r_0}\)
\( 1 + \epsilon \frac{\Delta r^2}{8r_0^2} \)
\\
&=
\frac{C\!\(k_\perp\)}{r_0^2}
\( 1 +\epsilon\,\frac{\Delta r^2}{8r_0^2} \)
\label{eq:limber_limit}\,,
\ea
with the flat-sky approximation $C(k_\perp)$.
Here, we assume that both $\Delta r/r$ and $|\epsilon|$ are small so that
$\ln(r+\Delta r/2)\simeq\ln(r)+\frac{\Delta r}{2r}$ and $r_0^\epsilon\simeq
1+\epsilon\ln(r_0)$. The last equality follows from the flat-sky
\cref{eq:kperp_to_infty_flat_sky} when $A_\mathrm{RSD}=1$ and the window is a
top-hat.

\cref{eq:limber_limit} clearly shows that the flat-sky approximation
has an intrinsic inaccuracy on small scales that is proportional to
the relative bin width $\Delta r^2/r^2$, and depends on the slope of the power
spectrum $-(3+\epsilon)$. This is the source of the discrepancy on small scales
between the exact calculation and the flat-sky calculation in the top-right
panel of \cref{fig:compare_fullsky_flatsky_sigma20}. 
This is somewhat complimentary to Limber's approximation which works
better for larger radial bins \cite{jeong/komatsu/jain:2009}.

Physically, the flat-sky discrepancy on small scales comes from
treating transverse separations for a given angle the same, whether they are
at the far end or the near end of the redshift bin. However, the ratio of
these transverse separations is
$(r_0 + \frac12\Delta r) / (r_0 - \frac12\Delta r) \approx 1 + \Delta r/r_0$
for small angles. Hence, the ratio $\Delta r/r_0$ appears in
\cref{eq:limber_limit}. The shape of the power spectrum also clearly matters,
as evaluating the power spectrum at smaller than center-of-bin scales at the
near end and larger-than-center-of-bin scales at the far end cancel only when 
$P(k) \propto k^{-3}$.
We, however, stress here that the difference stays sub-percent level even
for the pathological case ($\Delta r/r_0=0.4$) shown here.

The real-space comparison in \cref{fig:cl_exact_flat_limber} among the 
full calculation (blue solid lines), flat-sky approximation 
(orange dashed lines), and Limber approximation (Green dot-dashed lines)
clearly shows that the flat-sky approximation outperforms the Limber 
approximation. While the flat-sky and exact
calculations lie virtually on top of each other with percent-level
discrepancies (also see
\cref{fig:compare_fullsky_flatsky,fig:compare_fullsky_flatsky_sigma20}),
Limber's approximation does not approach the exact calculation until very 
large $\ell$. 

Incidentally, this large $\ell$ is also when the 
flat-sky approximation starts to break down.
\cref{fig:cl_flat_limber_ratio} compares the flat-sky and Limber's
approximation up until such high $\ell$ that the ratio becomes constant, and is
in rough agreement with \cref{eq:limber_limit} as well as the discrepancy 
shown in the top-right panel of \cref{fig:compare_fullsky_flatsky_sigma20}.

\section{RSD in Log-normal simulation}
\label{sec:comparisons}
\begin{figure*}
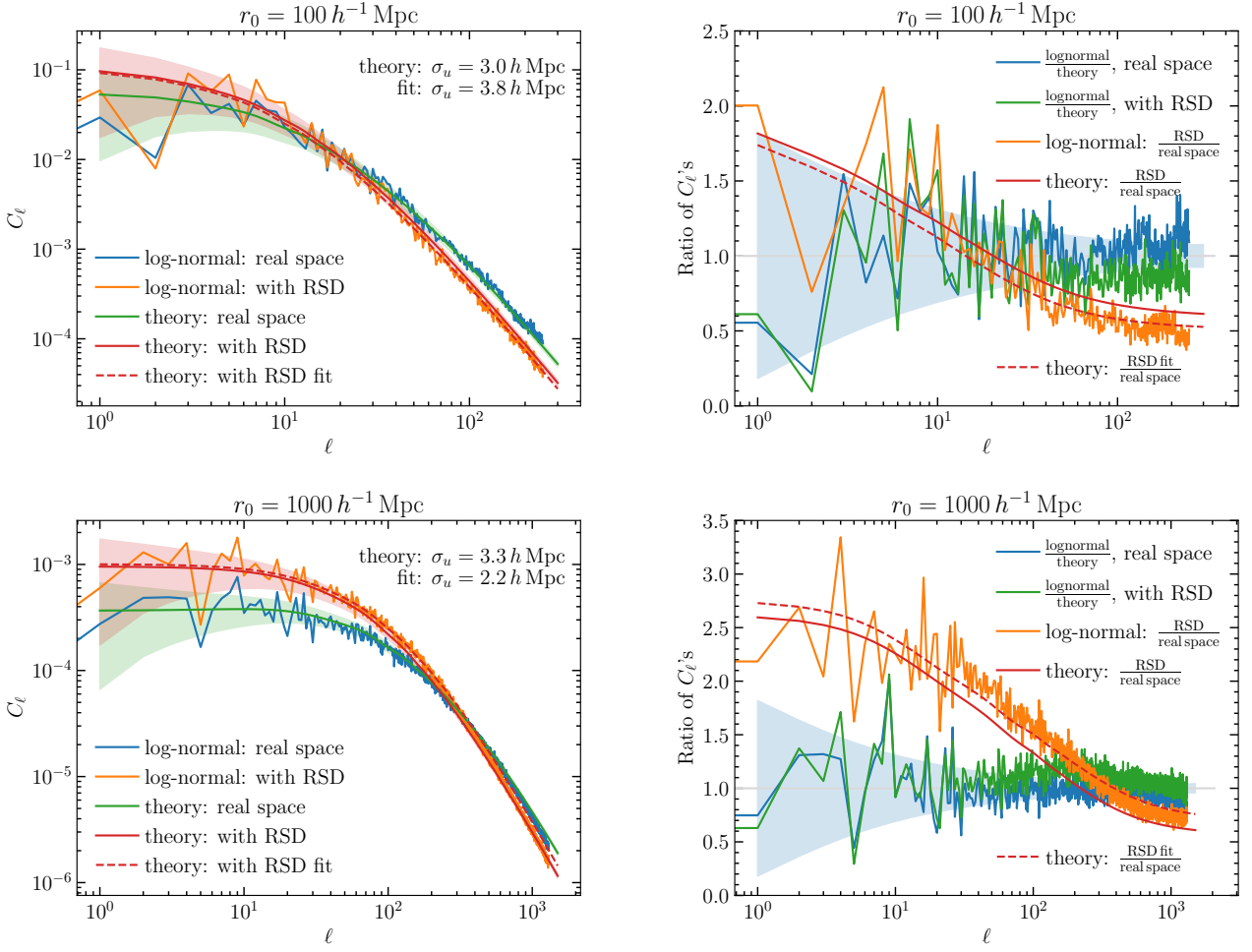

    \centering
    \incgraph{{figs/cl_lognormal_gaussian_r100.0_s4.0}.pdf}
    \incgraph{{figs/clratio_lognormal_gaussian_r100.0_s4.0}.pdf}
    \incgraph{{figs/cl_lognormal_gaussian_r1000.0_s4.0}.pdf}
    \incgraph{{figs/clratio_lognormal_gaussian_r1000.0_s4.0}.pdf}
    \caption{Comparison between theoretical calculation and log-normal
        simulations for a top-hat window function of width $\Delta
        r=\SI{8}{\per\h\mega\parsec}$, at $r_0=\SI{100}{\per\h\mega\parsec}$
        (top) and $r_0=\SI{1000}{\per\h\mega\parsec}$ (bottom).
        The left plots show the angular power spectrum with and without RSD,
        the left plots show the ratios between angular power spectra as
        indicated in the legend.
        For the solid theory lines we used \refeq{sigma_u} to calculate
        $\sigma_u$, for the dashed line we chose a value that leads to better
        match to the simulation result.
        The corresponding $k_\perp$-mode is obtained by
        $k_\perp=(\ell-\frac12)/r_0$.
    }
    \label{fig:lognormal}
\end{figure*}
Finally, in this section we compare the harmonic-space nonlinear RSD 
expression \refeq{Cl_RSD} with  the result from a log-normal simulation
\citep{Agrawal+:2017JCAP...10..003A}. 
Again, we adopt a top-hat window function of width $\Delta r=\SI{8}{\per\h\mega\parsec}$, and consider two radii of 
$r_0=\SI{100}{\per\h\mega\parsec}$ and $r_0=\SI{1000}{\per\h\mega\parsec}$.

For the $r_0=\SI{100}{\per\h\mega\parsec}$ simulation, 
we generate a cubic box with length $L_x=L_y=L_z=300$ and grid
size $N=600$ so that the resolution is \SI{0.5}{\per\h\mega\parsec}. We draw
\num{\sim2e6} galaxies. We then position the observer at the center of this
box, we shift the galaxies according to their line-of-sight velocity using
\ba
\label{eq:RSDshift}
s &= r + \frac{\vv\cdot\rhat}{aH}
\,,
\ea
where $\rhat$ is the line-of-sight unit vector. We then apply a top-hat
radial window function by limiting the sample to galaxies with redshift-space
distances $r_0-\frac12\Delta r\leq r\leq r_0+\frac12\Delta r$, where $\Delta
r=\SI{8}{\per\h\mega\parsec}$ and
$r_0=\SI{100}{\per\h\mega\parsec}$. This results in a sample
of $N_\mathrm{gal}=\num{7.7e5}$ galaxies in a spherical shell around the
observer. The angular power spectrum is measured from the simulation using the
\texttt{healpy}\footnote{\url{healpy.org}} software with
$N_\mathrm{side}=1024$ and distributing galaxies to their nearest grid point
on the sky. To measure the real-space angular power spectrum, we repeat this
without shifting the galaxies according to \refeq{RSDshift}.

For the second simulation we repeat this procedure with a cube of side length
$L=\SI{2160}{\per\h\mega\parsec}$, grid size $N=2160$, $n_\mathrm{side}=2048$,
and a total of \num{e9} galaxies. We then draw galaxies around
$r_0=\SI{1000}{\per\h\mega\parsec}$, leading to a sample of
$N_\mathrm{gal}=\num{9.9e6}$ galaxies in a shell around the observer.

We estimate the measurement uncertainty by 
\ba
\Delta C_\ell
&=
\sqrt{\frac{2}{2\ell+1}}\(C_\ell + \frac{4\pi}{N_\mathrm{gal}}\)\,,
\ea
but for the examples that we show here, the shot-noise contribution is 
negligibly small: that is what we have intended in order to test the 
RSD predictions on smaller scales.

In \reffig{lognormal}, we show the harmonic-space nonlinear RSD power 
spectrum from the log-normal simulations at low redshift ($r_0=100$ top panel) 
and high redshift ($r_0=1000$ bottom panel), along with corresponding 
theoretical predictions from \refeq{Cl_RSD}.
For both cases, the left panels show the power spectra for two cases
(1) without RSD (real space), and 
(2) with RSD (Kaiser effect + Gaussian FoG model).
To facilitate the comparison, we show various ratios of the angular 
power spectrum in the right panels: the ratio of the log-normal simulation 
to the theoretical calculation both in real space and in redshift space, 
and the ratio of redshift space to real space for
both the log-normal simulation and theoretical calculation.
For all cases, we find an excellent agreement between the simulation result and
the result from \refeq{Cl_RSD}.

For the solid lines in \reffig{lognormal}, we use the FoG model with 
the theoretical prediction for the one-dimensional velocity dispersion:
\ba
\label{eq:sigma_u}
\sigma_u^2
= \<s^2\> - \<s\>^2
= \frac{\<(\vv\cdot\rhat)^2\>}{a^2H^2}
&= \frac{f^2}{3}\int\frac{\dd^3k}{(2\pi)^3}\,\frac{P_m(k)}{k^2}\,,
\ea
where $P_m(k)$ is the matter power spectrum used as input to the
simulations. This results in the values indicated by ``theory'' in the
top-right corners of the panels on the left. We, however, find that we can
achieve a better match by fitting the velocity dispersion $\sigma_u$. The
values we chose are labeled ``fit'' in the figure, and the fitting results are
shown as the dashed lines.

\section{Conclusion}
\label{sec:conclusion}
In this paper, we present a novel method of calculating the harmonic-space
galaxy power spectrum including the nonlinear Kaiser effect. The general formula
in \refeq{Cl_RSD} states that nonlinear Kaiser effect can be modeled by modifying the radial window function.

We then apply the formula to model the nonlinear \emph{Fingers of God} effect
(FoG). We show that the FoG is equivalent to a smoothing of the radial window
function, and, unlike the three-dimensional RSD effect in Fourier space, the
FoG changes the harmonic-space power specturm on all scales. We considered
Gaussian, Lorentzian, and square-root-Lorentzian forms
[\refeqs{Afog_a}{Afog_c}] for the FoG. We show that for narrow window functions
the flat-sky approximation agrees with the wide-angle analysis within a few
tenths of a percent on all scales $\ell\geq2$ if we make the identification
$k_\perp r_0=\ell+0.5$. We also show that the flat-sky approximation
has a residual inaccuracy proportional to $(\Delta r/r)^2$ on all scales.
The flat-sky approximation, therefore, is most suitable for narrow radial bins,
and is complementary to Limber's approximation which is suitable for 
broader radial bins.

Comparing with the log-normal simulations shows an excellent agreement,
provided that the velocity dispersion parameter $\sigma_u^2$ is chosen to fit
the resulting power spectrum. The best-fitting $\sigma_u^2$ differs from the
measured variance in the line-of-sight pairwise velocity distribution function.

Note that the present paper only considers the auto-correlation with a thin 
redshift bin. As the flat-sky approximation has indicated, we are, 
therefore, primarily probing the clustering on the tangential directions, and
we lost radial correlation among different radial bins.
To fully exploit the three-dimensional galaxy distribution, it 
is therefore necessary to consider cross-correlations as well. 
\refeq{Cl_RSD} can also be used for such a task, and we
leave the details for a future investigation.

\acknowledgments
The authors thank Emanuele Castorina and Shun Saito for useful discussion.
This work was supported at Pennsylvania State University by NSF grant (AST-1517363) and NASA ATP program (80NSSC18K1103).

\bibliography{references}

\end{document}